\documentclass[superscriptaddress,twoside,twocolumn,showpacs,preprintnumbers,amssymb,pre]{revtex4}
\usepackage{graphicx}
\usepackage{dcolumn}
\usepackage{bm}
\begin{document}
\title{Effects of Buckling on Stress and Strain in Thin
       Randomly Disordered Tension-Loaded Sheets}
\author{Bj{\o}rn Skjetne}
 \affiliation{Department of Chemical Engineering,
              Norwegian University of Science and Technology,\\
              N-7491 Trondheim, Norway}
 \affiliation{Department of Physics,
              Norwegian University of Science and Technology,\\
              N-7491 Trondheim, Norway}
\author{Torbj{\o}rn Helle}
 \affiliation{Department of Chemical Engineering,
              Norwegian University of Science and Technology,\\
              N-7491 Trondheim, Norway}
\author{Alex Hansen}
 \affiliation{Department of Physics,
              Norwegian University of Science and Technology,\\
              N-7491 Trondheim, Norway}
\date{\today}
\begin{abstract}
We study how crack buckling affects stress and strain in a thin 
sheet with random disorder. The sheet is modeled as an elastic 
lattice of beams where each of the beams have individual 
thresholds for breaking. A statistical distribution with an 
exponential tail towards either weak or strong beams is used 
to generate the thresholds and the magnitude of the 
disorder can be varied arbitrarily between zero and infinity. 
Applying a uniaxial force couple along the top and bottom rows of 
the lattice, fracture proceeds according to where the ratio of 
the stress field to the local strength is most intense. Since 
breakdown is initiated from an intact sheet where the first crack 
appears at random, the onset and mode of buckling varies 
according to where and how the cracks grow. For a wide range of 
disorders the stress-strain relationships for buckling sheets 
are compared with those for non-buckling sheets. The ratio of 
the buckling to the non-buckling value of the maximum external 
force the system can tolerate before breaking is found 
to decrease with increasing disorder, as is the ratio for the 
corresponding displacement.
\end{abstract}
\pacs{81.40.Jj, 62.20.-x, 05.40.-a}
\maketitle

\section{Introduction}
\label{intro}
In recent years methods have been developed within the 
statistical physics community to describe breakdown phenomena in 
complex media~\cite{smod}. These are the so-called lattice 
models, where the material is reduced to a set of points on
a grid whereupon disorder is imposed on each of the elements 
on the grid.
The desire to understand structurally non-uniform systems 
stems from the fact that many materials, natural or man-made, 
show a significant degree of disorder on the microscopic or 
mesoscopic level. In order to realistically describe how such 
materials fracture one has to include the interplay between, 
on the one hand, local variations in material properties and, 
on the other hand, a constantly evolving non-uniform stress 
field. The above mentioned lattice models are especially 
well suited for this purpose.

Most of the work done with lattice models on fracture 
and other breakdown phenomena, however, has focused on the fundamental 
underlying principles rather than traditional problems in 
fracture mechanics. The various quantities studied have been 
expressed through scaling laws and critical exponents, 
often with the aim to shed light on universal aspects of 
phenomena which are seemingly unrelated. The most common
examples besides fracture
are transport properties and growth processes~\cite{bara,bouc}. 
Obviously there is much to benefit from the application 
of lattice modeling to more specific problems in fracture 
mechanics, especially where disordered materials are concerned.

By far the most popular tool in fundamental studies of breakdown 
processes has been the so-called random fuse model~\cite{fuse}, 
a scalar analogue of fracture which really 
models electrical breakdown. Another model, which takes account 
of the vectorial nature of elasticity, is the beam 
lattice~\cite{roux,herm}. Recently, we introduced a 
three dimensional version of the beam lattice which is 
suitable to describe buckling in thin planar 
structures~\cite{tbp1}. Such buckling behaviour
is perhaps most frequently associated with thin 
plates or beams under compressive loading. In this paper we 
concern ourselves with the special case of a thin planar 
structure under tensile, mode-I type, loading. 
The interaction 
of buckling with fracture in such cases is a well known
phenomenon, although as a problem it remains much less 
studied.

Most of the data reported, both theoretical and experimental, 
have centered on a few, rather limited, special cases, 
such as that of a thin plate with a center-crack,
aligned in a perpendicular fashion to the 
externally applied force.
When such a plate is subjected to uniaxial tensile 
loading, transverse compressive stresses build up in the 
vicinity of the crack, causing the unsupported edges to 
deflect out of the initial rest plane. This redistributes
the stresses around the crack and leads to a stronger 
singularity at the tip, thus reducing the external force
necessary to propagate crack growth. 
There are many applications for which the special case
of a homogeneous thin plate with a center-crack is 
representative. Crack buckling, however, is observed under 
a variety of conditions, and often involves anisotropic 
or disordered materials with more than one crack. 

Composites, for instance, are on the increase as the 
preferred material for use in the thin walled plate- or 
shell-structures so essential to the construction of vehicles 
for transportation purposes, e.g., hulls and fuselages. 
Critical loads for orthotropic plates have been obtained 
in finite element (FEM) calculations, but only within the usual 
single-crack or hole scenario~\cite{lars}. The importance of buckling 
and the way it interacts with fracture in the presence of 
multiple cracks has been recognized for some time, however. 
In the aerospace industry, for instance, one seeks to make 
allowance in the design approach for the presence of 
multi-site damage, i.e., assess to what extent a series of 
aligned cracks have on the strength properties of a 
structure~\cite{sesh,weli}. Moreover, buckling plays 
an important role in the breaking of thin sheets where the
disordered nature of the micro-structure cannot be
ignored. The crack geometry which obtains in such cases 
may be highly complex. A specific example of this is 
paper. Paper results from a rapid filtration process 
involving water and wood fibres. The sheet formed is a 
layered fibre-structure, but nonetheless strongly coupled 
in the vertical direction. Paper is thus a highly stochastic 
material, where the essentially random structure is 
modified by flocculation, i.e., an undesired clustering 
of fibres in the early stages of the filtration 
process~\cite{dods}. With conditions of tensile loading 
frequently arising in production facilities as well as 
printing presses, buckling deformations due to tension in 
paper is a well known phenomenon~\cite{ybse}. Its 
interaction with fracture has not received sufficient
attention, however. 

In the following we briefly summarize some of the research 
that has been done on the buckling of thin sheet materials
under tensile loading. Among the earliest investigations was 
that made by Cherepanov~\cite{cher} on membranes containing 
holes. This, and much of the literature which followed mainly
concerned itself with the calculation of critical loads for 
the onset of buckling~\cite{many,mark,gcsi,shaw}, rather than 
the effect buckling has on the fracture properties once it 
has set in, i.e., the so-called post-buckling behaviour. 
That buckling should adversely affect residual strength has 
been recognized for some time, however, with early experimental 
observations reported by Forman~\cite{form}, Dixon and 
Strannigan~\cite{dixo}, and Zielsdorff and Carlson~\cite{ziel}, 
hence the interest in determining the loads and conditions 
under which plates with specified parameters buckle. As
already noted, the source to this reduction in strength has 
been traced to a redistribution of stresses which leads to 
a stronger singularity at the tip of the crack~\cite{pety,riks,baru}. 

Most of the results relevant to the critical 
buckling load have been obtained for thin plates with a
center crack. Such results are usually
expressed in the form of an empirical 
relation which involves plate thickness, crack length, Young's
modulus and a proportionality factor. In their recent 
experimental work, Guz and Dyshel have also considered several 
cases which can be seen as variations on the theme of a 
central crack; e.g., the effect that crack curvature or an 
inclination angle has on either the critical buckling load or the
residual strength of a plate with a centrally located 
crack~\cite{guz1}; or the effect a straight central crack 
has on the critical buckling load of a two-layered 
plate~\cite{guz2}. Centrally cracked plates are not the only
systems studied, however, plates with edge cracks have also been
considered. Here the buckling 
mechanism has been found to be different from that which 
causes a central crack to bulge~\cite{carl}. Critical buckling 
loads relevant to both perpendicular~\cite{guz3,dysh} and 
inclined~\cite{dys2} edge cracks have been obtained, as well 
as results for the effect buckling has on the residual 
strength of edge cracked panels~\cite{dysh}. 

With regard to modeling and theoretical research, an early 
study by Pellet et al. employed a Rayleigh-Ritz variational 
procedure to obtain critical buckling loads in the presence 
of a circular hole~\cite{pell}. Recent FEM calculations 
realistically reproduce the observed buckling behaviour 
around centrally located cracks, and results have been 
obtained for critical loads which agree well with experimental 
findings~\cite{mark,shaw,pety,riks,baru}. The FEM approach 
has also been used to study the various modes of buckling and 
the extent of the buckling zone, e.g., for plates with 
either a perpendicular~\cite{pety} or an inclined~\cite{baru}
central crack. Gilabert et al. also obtained results 
relevant to the zone of deformation~\cite{gila}, 
and critical loads for various crack geometries,
e.g., circular holes or rectangular cut-outs with
sharp or rounded corners, have been obtained in other FEM
calculations~\cite{shim}.

Features of the post-buckling behaviour, other than the 
shape and extent of the buckling zone, was obtained by Petyt, 
i.e., for the vibration characteristics of a centrally 
cracked plate subject to acoustic loads~\cite{pety}. Petyt 
also addressed the non-linear nature of FEM calculations 
for the post-buckling behaviour, and Riks et al.~\cite{riks} 
used such an analysis to show that the energy release-rate 
at the tip of the crack undergoes a sudden increase at the 
onset of buckling. The stress intensity along the 
post-buckling path is then larger than that obtained along 
the pre-buckling path for the same load, a result which, in 
agreement with experimental observations, indicates that the 
residual strength of the plate is reduced by buckling. The 
effect of crack inclination on the energy release-rate in 
the post-buckling state has also been studied~\cite{baru}. 
FEM calculations for the load versus crack-opening length in 
buckling and non-buckling fracture modes have been 
carried out in a study by Seshadri and Newman~\cite{sesh}, 
showing a significant reduction in the residual strength. 
Their work also considered the effect of plasticity by 
assuming a hypothetical material with a very high crack-tip 
opening angle, with the reduction in strength due to buckling
now being generally less pronounced than in the brittle case.

As the above summary shows,
practically all previous work considers the effect
buckling has on the strength properties of an already
cracked plate, or a plate with a geometrical discontinuity 
such as a circular hole or a rectangular cut-out. 
In other words, if the 
physical parameters of the plate are such that buckling 
can be expected before the crack begins to grow, the
residual strength of the plate will be significantly
lower than what would otherwise be expected, based on an
analysis which does not take account of buckling. The
present study of fracture and buckling is fundamentally
different in the sense that we regard a sheet which, in 
its initial state, has no cracks or other discontinuities.
Instead, cracks form by a complex process which depends 
on the evolving distribution of stresses 
and its interaction with a disordered meso-structure.
The onset of buckling in this scenario, and the effect
buckling has on the fracture properties, will vary according
to the type of disorder used, i.e., weak or strong. 
Whereas for strong disorders there will be significant
sample-to-sample variations, such variations
tend to be less pronounced for weak disorders. However,
even for weak disorders the final crack
which breaks the system will only rarely appear at the exact 
center of the sheet, and even then the situation might
be complicated by additional cracks in the vicinity -- cracks 
which interact with the main crack so as to alter the 
distribution of stresses and hence also the exact shape or 
mode of buckling. Therefore, due to the statistical nature 
of the results obtained, features such as the extent of the 
buckling zone, or the shape of the deflected crack edge, 
will not at present be dealt with in any detail. For the 
same reasons critical loads are not calculated, since the 
magnitude of this quantity depends on very specific sheet 
parameters, i.e., for a given sheet thickness the critical
load has been shown to depend on the ratio of the crack 
length to the sheet width.

\section{The Beam Lattice}
\label{planb}
The beam lattice used in our calculations is a regular
square lattice, where each beam has unit length. System
size $L$ therefore corresponds to the number of beams
along the top or bottom rows. The nodes are equidistantly
spaced along $J=L+2$ horizontal rows and $I=L+1$ vertical
columns, each having four nearest neighbours to
which it is fastened by elastic beams. When nodes
are displaced the angle at the joint where two beams 
come together remains perpendicular, thus
inducing shearing forces and bending moments in addition
to axial tension or compression.

In the plane beam lattice there are three degrees of 
freedom for the displacement of nodes, i.e., translation 
along either the $X$-axis or the $Y$-axis, and rotations
about the $Z$-axis. The displacement field is obtained by
requiring the sum of forces and moments on each node to be
zero. Specifically, we solve
\begin{eqnarray}
    \sum_{j}D_{ij}
             \left[\begin{array}{l}
                      x_{i}\\
                      y_{i}\\
                      w_{i}\\
                    \end{array}
             \right]=\lambda
             \left[\begin{array}{l}
                      X_{i}\\
                      Y_{i}\\
                      W_{i}\\
                   \end{array}
             \right],
              \label{ma2x}
\end{eqnarray}
where the forces on node $i$ are
\begin{eqnarray}
    X_{i}&=&{_{x}A_{i}^{(1)}}+{_{x}T_{i}^{(2)}}
         +{_{x}A_{i}^{(3)}}+{_{x}T_{i}^{(4)}},
         \label{sumfx}\\
    Y_{i}&=&{_{y}T_{i}^{(1)}}+{_{y}A_{i}^{(2)}}
         +{_{y}T_{i}^{(3)}}+{_{y}A_{i}^{(4)}},
         \label{sumfy}\\
    W_{i}&=&\sum_{j=1}^{4}{_{w}M_{i}^{(j)}},
         \label{sumfw}
\end{eqnarray}
by numerical relaxation, i.e., the 
conjugate gradient method~\cite{hest}, to obtain the set of 
displacements which minimizes the elastic energy of the 
lattice.

In Eqs~(\ref{sumfx}) and (\ref{sumfy}), $A$ and $T$ denote
axial and transverse force, respectively, while in
Eq.~(\ref{sumfw}) $M$ denotes the bending moment. Hence,
${_{x}A_{i}^{(3)}}$ is the force exerted on node $i$ 
from $j=3$ along the $X$-axis by axial tension or compression. 
Neighbouring nodes are numbered 
anti-clockwise, starting with $j=1$ on the left. 

Defining $\delta r=r_{j}-r_{i}$, where $r\in\{x,y,w\}$, the 
contributions from $j=1$ are
\begin{eqnarray}
    _{x}A_{i}^{(1)}\hspace{-2mm}&=&\hspace{-2mm}
                \frac{1}{\alpha}\delta x,
                 \label{fi}\\
    _{y}T_{i}^{(1)}\hspace{-2mm}&=&\hspace{-2mm}
                \frac{1}{\beta+\frac{\gamma}{12}}
                 \bigl[\delta y-\frac{1}{2}\bigl(w_{i}+w_{j}
                  \bigr)\bigr],
                   \label{si}\\
    _{w}M_{i}^{(1)}\hspace{-2mm}&=&\hspace{-2mm}
                \frac{1}{\beta+\frac{\gamma}{12}}
                 \bigl[\hspace{0.5mm}\frac{\beta}{\gamma}
                  \delta w
                  +\frac{\delta y}{2}
                   -\frac{1}{3}(w_{i}
                    +\frac{w_{j}}{2})
                      \hspace{0.5mm}\bigr],
                       \label{mi}
\end{eqnarray}
where 
\begin{eqnarray}
    \alpha=\frac{1}{E\rho},\hspace{3mm}
     \beta=\frac{1}{G\rho},\hspace{3mm}
      \gamma=\frac{1}{EI},
       \label{mate}
\end{eqnarray}
are the prefactors characteristic of the material 
and its dimensions, i.e., $E$ is Young's modulus, $\rho$ and $I$ 
the area of the beam section and its moment of inertia, 
respectively, and $G$ the shear modulus~\cite{herm}.

The fracture process consists of removing one beam at a time,
whereby 
a new set of displacements are obtained at each step 
by solving Eq.~(\ref{ma2x}). The criterion by
which a beam is removed from the lattice depends on the ratio
of the local stress to the breaking threshold. Using
$t_{A}$ and $t_{M}$ for the maximum thresholds in axial force
and bending moment, respectively, a good breaking 
criterion~\cite{herm} inspired from Tresca's theory is 
\begin{equation}
    \left(\frac{A}{t_{A}}\right)^{2}+
     \frac{|M|}{t_{M}}\geq 1,
      \label{tresca}
\end{equation} 
where $|M|={\rm max}(|M_{i}^{(j)}|,|M_{j}^{(i)}|)$ is the 
largest of the moments at the two beam ends $i$ and $j$. 

The time taken for mechanical equilibrium to be reached
is assumed to be much shorter than the time taken to remove
a beam, i.e., the fracture process is assumed to be
quasi-static. It is
driven by imposing a fixed unit 
displacement on the top row of the lattice. Since internal 
displacements, forces and moments are proportional to this, 
the actual external elongation of the lattice is obtained by
determining the minimum value of the proportionality
constant $\lambda_{L}$ in
\begin{equation}
    \left(\lambda_{L}\frac{A}{t_{A}}\right)^{2}+
     \lambda_{L}\frac{|M|}{t_{M}}=1,
      \label{2tresca}
\end{equation} 
from which the external force is obtained as
\begin{eqnarray}
    f_{L}=\lambda_{L}\hspace{-1mm}
           \sum_{i=1}^{I(J-1)}
            \hspace{-1mm}{_{y}A_{i}^{(2)}}n_{i,y},
             \label{extfor}
\end{eqnarray}
with the array
\begin{eqnarray}
    n_{i,y}=\left\{
             \begin{array}{c}
               1,\\
               0,
             \end{array}
            \right.
             \label{niy}
\end{eqnarray}
keeping track of whether beams are intact (1) or have been
broken (0), respectively. 

In Eq.~(\ref{extfor}), contributions other than 
${_{y}A_{i}^{(2)}}$ cancel when the sum is over the entire
lattice. This is due to the square lattice topology
and the nature of the external 
boundary conditions applied, i.e., mode-I type loading in 
the $Y$-direction.
Regarding the internal forces, the
same term continues to be the sole non-zero contribution
when the lattice is intact. Consequently the first 
beam to break is that for which the ratio $A/t_{\rm A}$ is 
the largest. After this has been removed, however, bending 
moments $M$ and transverse forces $T$ (including shear) are 
induced in the immediate neighbourhood of the beam. This is 
due to the screening effect of the ``hole'', or crack, 
created by its removal from the lattice. 

\begin{figure} [t]
\includegraphics[angle=-90,scale=0.85]{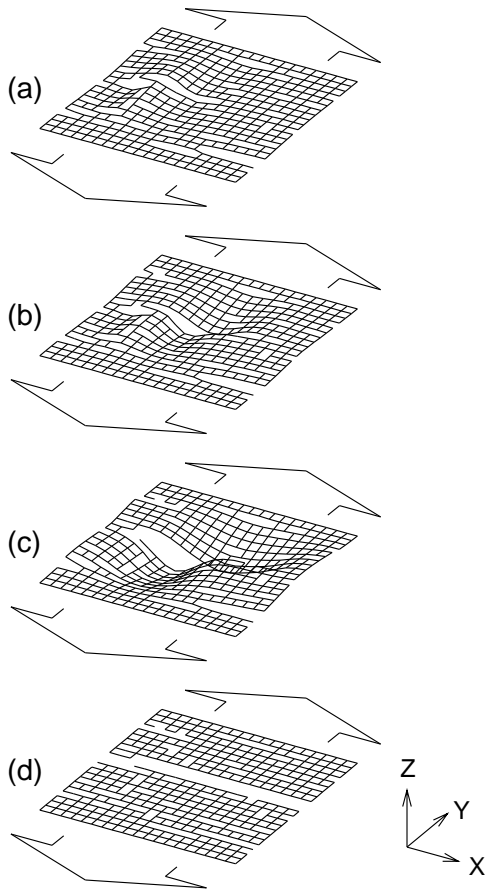}
\caption{A disordered lattice of size $L=20$, 
         shown at four different stages in the breakdown
         process. The lattice is strained to failure by 
         applying a force couple at the top and bottom. 
         With the appearance of large cracks, the structure
         is seen to deflect out of the initial rest plane.
         The number of broken beams in this simulation 
         are, from~(a) to~(d), $N=65$, 72, 78, and 86, 
         respectively.
         \label{warp}}
\end{figure}

A thin sheet will usually display deviations in symmetry with
respect to the thickness, e.g., there may be variations in the
thickness itself or there may be a gradient in the structural 
properties of the material. An example of the latter is paper,
where, due to the process by which it is manufactured, the 
fibre structure on one side always has a stronger orientational 
bias. In other materials the density varies in the thickness 
direction. When a uniaxial force couple is applied on opposite 
edges of the sheet, such variations create bending moments 
about axes in parallel within the $XY$-plane, see 
Fig.~\ref{warp}, which shows the coordinate system and the 
direction of the external load. In fact, when the 
internal stresses (which arise as a consequence of the external 
load condition) combine with certain crack configurations,
minute deviations of the symmetry plane itself from a perfect 
two dimensional embedding will be sufficient to cause
buckling. Numerous studies have been reported 
in the literature concerning the external load necessary to
cause buckling, e.g., for a sheet with a central crack
the magnitude of the critical load has been found to decrease 
with decreasing 
sheet thickness and increasing crack extent.

The additional terms which cause buckling 
are much smaller in magnitude than those governing the forces
within the plane lattice. There is, however, a non-separable
relationship between in-plane and out-of-plane displacements
which causes the in-plane coordinates of the non-buckling 
lattice to change significantly when buckling is allowed. For
this reason the $X_{i}$, $Y_{i}$ and $W_{i}$ components of
the buckling lattice~\cite{tbp1}
\begin{eqnarray}
    \sum_{j}D_{ij}
             \left[\begin{array}{l}
                      u_{i}\\
                      v_{i}\\
                      w_{i}\\
                      x_{i}\\
                      y_{i}\\
                      z_{i}\\
                    \end{array}
             \right]=\lambda
             \left[\begin{array}{l}
                      U_{i}\\
                      V_{i}\\
                      W_{i}\\
                      X_{i}\\
                      Y_{i}\\
                      Z_{i}\\
                   \end{array}
             \right]
              \label{ma3x}
\end{eqnarray}
contain additional non-linear terms, i.e., terms not included in
Eq.~(\ref{ma2x}). 

Specifically, the axial force component
\begin{eqnarray}
    X_{i}^{(1)}=\frac{1}{\alpha}\delta x
                 \label{oldxif}
\end{eqnarray}
in Eq.~(\ref{ma2x}) is replaced by
\begin{widetext}
\begin{eqnarray}
    X_{i}^{(1)}
      &=&-F_{i}^{(1)}\left[\cos{w_{i}}\cos{u_{i}}-
          \frac{\delta w}{2}\cos{\delta w}\sin{w_{i}}
          -\frac{\delta u}{2}\cos{\delta u}\sin{u_{i}}\right]
            \label{xif}
             \\
      &+&\frac{1}{\beta+\frac{\gamma_{\rm Z}}{12}}
          \left[\bigl(1+\delta x\bigr)\sin{w_{i}}-
           \delta y\cos{w_{i}}+\frac{\delta w}{2}\right]\sin{w_{i}}
           +\frac{1}{\beta+\frac{\gamma_{\rm Y}}{12}}
             \left[\bigl(1+\delta x\bigr)\sin{u_{i}}-
              \delta z\cos{u_{i}}+\frac{\delta u}{2}\right]
               \sin{u_{i}},
                \nonumber
\end{eqnarray}
where
\begin{eqnarray}
    F_{i}^{(1)}=\frac{1}{\alpha}
      \left\{1-
       \frac{\delta u}{2}\bigl[sin(\frac{\delta u}{2})\bigr]^{-1}
        \sqrt{\delta z^{2}+\bigl( 1+\delta x\bigr)^{2}}
         \hspace{1mm}\right\}
          \label{fiforce}
\end{eqnarray}
is the force along the axis of the beam, including an
angular correction which takes into account the additional
elongation due to bending. Likewise, the transverse force
\begin{eqnarray}
    Y_{i}^{(1)}=
      \frac{1}{\beta+\frac{\gamma}{12}}
       \Bigl[\delta y-\frac{1}{2}\bigl(w_{i}+w_{j}
        \bigr)\Bigr]
         \label{oldyif}
\end{eqnarray}
is replaced by
\begin{eqnarray}
    Y_{i}^{(1)}=-F_{i}^{(1)}
         \left(\sin{w_{i}}\cos{u_{i}}-
          \frac{\delta w}{2}\cos{\delta w}\cos{w_{i}}
           \right)
           -\frac{1}{\beta+\frac{\gamma_{\rm Z}}{12}}
             \left[\bigl(1+\delta x\bigr)\sin{w_{i}}-
              \delta y\cos{w_{i}}+\frac{\delta w}{2}\right]\cos{w_{i}},
               \label{yif}
\end{eqnarray}
and the in-plane moment
\begin{eqnarray}
    W_{i}^{(1)}=\frac{1}{\beta+\frac{\gamma}{12}}
                 \left[\hspace{0.5mm}\frac{\beta}{\gamma}
                  \delta w
                  +\frac{\delta y}{2}
                   -\frac{1}{3}\Bigl(w_{i}
                    +\frac{w_{j}}{2}\Bigr)
                      \hspace{0.5mm}\right]
                       \label{oldwif}
\end{eqnarray}
is replaced by
\begin{eqnarray}
    W_{i}^{(1)}=\frac{\beta}{\gamma_{\rm Z}(\beta+
                 \frac{\gamma_{\rm Z}}{12})}\delta w
                 -\frac{1}{2(\beta+\frac{\gamma_{\rm Z}}{12})}
                   \left[
                    \bigl(1+\delta x\bigr)\sin{w_{i}}-
                     \delta y\cos{w_{i}}+\frac{\delta w}{3}
                      \right]-F_{i}^{(1)}
                       \frac{\delta w}{4}\cos{\delta w}.
                        \label{wif}
\end{eqnarray}

The additional terms of Eq.~(\ref{ma3x}) are
\begin{eqnarray}
    Z_{i}^{(1)}=-F_{i}^{(1)}
             \left(\sin{u_{i}}+
              \frac{\delta u}{2}\cos{\delta u}\cos{u_{i}}
               \right)
               -\frac{1}{\beta+\frac{\gamma_{\rm Y}}{12}}
                 \left[\bigl(1+\delta x\bigr)\sin{u_{i}}-
                  \delta z\cos{u_{i}}+\frac{\delta u}{2}
                   \right]\cos{u_{i}}
                    \label{zif}
\end{eqnarray}
for the displacements normal to the $XY$-plane, and
\begin{eqnarray}
    U_{i}^{(1)}=
      \frac{\beta}{\gamma_{\rm Y}(\beta+\frac{\gamma_{\rm Y}}{12})}
       \delta u-\frac{1}{2(\beta+\frac{\gamma_{\rm Y}}{12})}
        \left[\bigl(1+\delta x)\sin{u_{i}}-
         \delta z\cos{u_{i}}+
          \frac{\delta u}{3}\right]-F_{i}^{(1)}
           \frac{\delta u}{4}\cos{\delta u}
            \label{uif}
\end{eqnarray}
for the rotations about the $Y$-axis. Finally,
\end{widetext}
\begin{eqnarray}
    V_{i}^{(1)}=\xi\delta v
                 \label{vif}
\end{eqnarray}
is the torque of the beam when rotations are about the
$X$-axis. Assuming $w>t$, 
\begin{eqnarray}
    \xi=G\frac{wt^{3}}{3}
          \label{eqxi}
\end{eqnarray}
is the torsional moment of inertia in Eq.~(\ref{vif}), with $w$ 
denoting the width of the beam cross section and $t$ its 
thickness. 
Assuming a rectangular cross-section,
the moments of inertia for bending are
\begin{eqnarray}
    I_{\rm Z}=\frac{1}{12}w^{3}t
\end{eqnarray}
within the $XY$-plane, and
\begin{eqnarray}
    I_{\rm X}=\frac{1}{12}wt^{3}=I_{\rm Y}
\end{eqnarray}
within the $YZ$- and $XZ$-planes, respectively.

The expressions for the forces acting on the beams in 
Eqs.~(\ref{xif}) to~(\ref{vif}) have been
derived by considering an elastic beam with no end 
restraints~\cite{roar}, where the ratio of the beam width to
the thickness presently has been set to 10:1. With regard 
to bending flexibility, the lattice is now more pliable in the 
out-of-plane direction, as would be expected for a thin sheet 
material.

In lattice modeling the rule by which a beam is broken can be
specified according to the properties of the material one
wishes to study. Presently the fracture criterion is taken
to depend on a combination of axial stress, bending and
torsion. Hence, we assume
\begin{equation}
    \left(\frac{F_{\rm C}}{t_{F_{\rm C}}}\right)^{2}+
     \frac{|\mu_{\rm C}|}{t_{\mu_{\rm C}}}\geq 1,
      \label{bcri}
\end{equation} 
where 
\begin{eqnarray}
    F_{\rm C}=F_{i}^{(j)}-\chi
        \Bigl|Q_{i}^{(j)}        
         \Bigr|                 
          \label{ft}
\end{eqnarray}
is the effective stress, and
\begin{eqnarray}
    Q_{i}^{(j)}=\sum_{k=1,3}\delta_{kj}V_{i}^{(j)}
               +\sum_{k=2,4}\delta_{kj}U_{i}^{(j)}
                 \label{qq}
\end{eqnarray}
is the torque. Moreover, with $\mu_{\rm C}$ denoting the combined 
bending moment and
\begin{eqnarray}
    \sigma=\frac{w}{t}
\end{eqnarray}
being the aspect ratio of the cross section of the beam, the
expression
\begin{equation}
    \chi=\left\{
         \begin{array}{cl}
            1+(\sigma L)^{2}\Bigl|F_{i}^{(j)}\Bigr|, 
                    & \hspace{2mm}F_{i}^{(j)}<0,\\
            1,                                           
                    & \hspace{2mm}F_{i}^{(j)}\geq0,
         \end{array}\right.
          \label{tsens}
\end{equation}
is an enhancement factor in Eq.~(\ref{ft}). In Eq.~(\ref{qq})
the Kronecker delta has been used to distinguish between the
four neighbouring beams.

Angular displacements about the $X$- or $Y$-axis in 
Eq.~(\ref{bcri}) activate the stress enhancement mechanism. This 
increases the stress in a beam when it is under axial tensile
loading. Specifically, the larger the the load is, the more 
sensitive the beam will be to the presence of a certain amount
of axial torque. Compressive loads are assumed to be less
important, with torque now instead removing some of the axial 
compression. 

The exact mechanism by which buckling alters the rupture mode
of a thin sheet will probably vary according to material
properties, structural composition, and so on. In many cases
it is a well known fact that the work required to drive
a crack across a given area is much smaller in mode-III tearing
than in pure mode-I tension, as is easily verified with a
piece of paper. Transverse forces, however, are not presently 
assumed to contribute. This is because the disorder of the 
sheet is modeled on a mesoscopic scale. In a material such 
as paper, tearing is a shear displacement which affects 
material properties on much smaller scales, e.g., on the level 
of the individual fibres. In the beam lattice, on the other
hand, each individual beam is representative of the sheet on 
the level of fibre flocculations. The effect of mode-III crack 
propagation is instead included by the above combination of 
torsion and axial stress.

\section{Numerical Scheme of Calculations}
\label{numscheme}
Mathematically, conjugate gradients is an iterative method to
obtain the minimum of a quadratic expression, in our case
the elastic energy. For the energy to be quadratic, however,
the forces involved must be linear. In obtaining a numerical 
solution, therefore, the presence of non-linear terms is a 
complicating factor. Nonetheless, provided the proper numerical 
safeguards are employed, the correct minimum can be found 
effectively by use of conjugate gradients. Specifically, 
\begin{eqnarray}
    \sum_{j=1}^{4}X_{i}^{(j)}=
    \sum_{j=1}^{4}Y_{i}^{(j)}=
    \sum_{j=1}^{4}W_{i}^{(j)}=0
    \label{sumxyw}
\end{eqnarray}
is the solution obtained by relaxing the in-plane coordinates
while keeping the out-of-plane coordinates fixed. Since the 
leading terms of $X_{i}$, $Y_{i}$ and $W_{i}$ are all linear, 
the actual solution for this in-plane projection always lies 
close to its linear solution. It is found by re-initializing the 
search, each time using conjugate gradients starting from the 
previous linear approximation. This is repeated until the 
minimum stops changing, typically 6-7 searches are required, 
with convergence being rapid. 

After this intermediate solution has 
been obtained, it is frozen, whereupon the out-of-plane 
coordinates are relaxed, one at a time. In this case, however, 
the leading terms are non-linear. Consequently a single
search is made toward the minimum to obtain
\begin{eqnarray}
    \sum_{j=1}^{4}Z_{i}^{(j)}\approx 0,
    \label{sumz}
\end{eqnarray}
i.e., a partial only, or (at best) very approximate, solution. 
Moreover, in order to ensure that this incomplete move carries 
towards (and not away from) the minimum, the step size of the 
conjugate gradient iterations in this phase is reduced to a 
much smaller value. 

Likewise, the out-of-plane angular displacements are updated, 
one at a time, using the same down-scaled step size, to
obtain
\begin{eqnarray}
    \sum_{j=1}^{4}U_{i}^{(j)}\approx 0,\hspace{3mm}
    \sum_{j=1}^{4}V_{i}^{(j)}\approx 0,
    \label{sumuv}
\end{eqnarray}
while keeping all other coordinates fixed. 

After re-setting the iterational step size, the whole 
procedure outlined above is repeated. The updated 
coordinates obtained for the out-of-plane displacements, 
approximate as they are, do nonetheless cause the in-plane 
displacements to change. As the final buckled configuration 
of the lattice is approached the quality of the 
intermediate partial solutions, represented by Eqs.~(\ref{sumz}) 
and~(\ref{sumuv}), gradually improves. 

Hence, after a number of repetitions we obtain
\begin{eqnarray}
    \sum_{j=1}^{4}X_{i}^{(j)}=
    \sum_{j=1}^{4}Y_{i}^{(j)}=... =
    \sum_{j=1}^{4}V_{i}^{(j)}=0,
    \label{sumall}
\end{eqnarray}
for the sum of forces and moments on all nodes. The
previous set of displacements is now identical to the current 
set of displacements, the calculation having converged upon 
the final solution. 

\section{Disorder}
\label{dx}
Each time a beam is broken, a new set of displacements is
calculated according to the scheme outlined in 
section~\ref{numscheme}. A fundamental factor deciding
how the lattice breaks is the choice made for the type and
magnitude of 
disorder in the distribution of breaking thresholds.
\begin{figure*}
\includegraphics[angle=-90,scale=0.62]{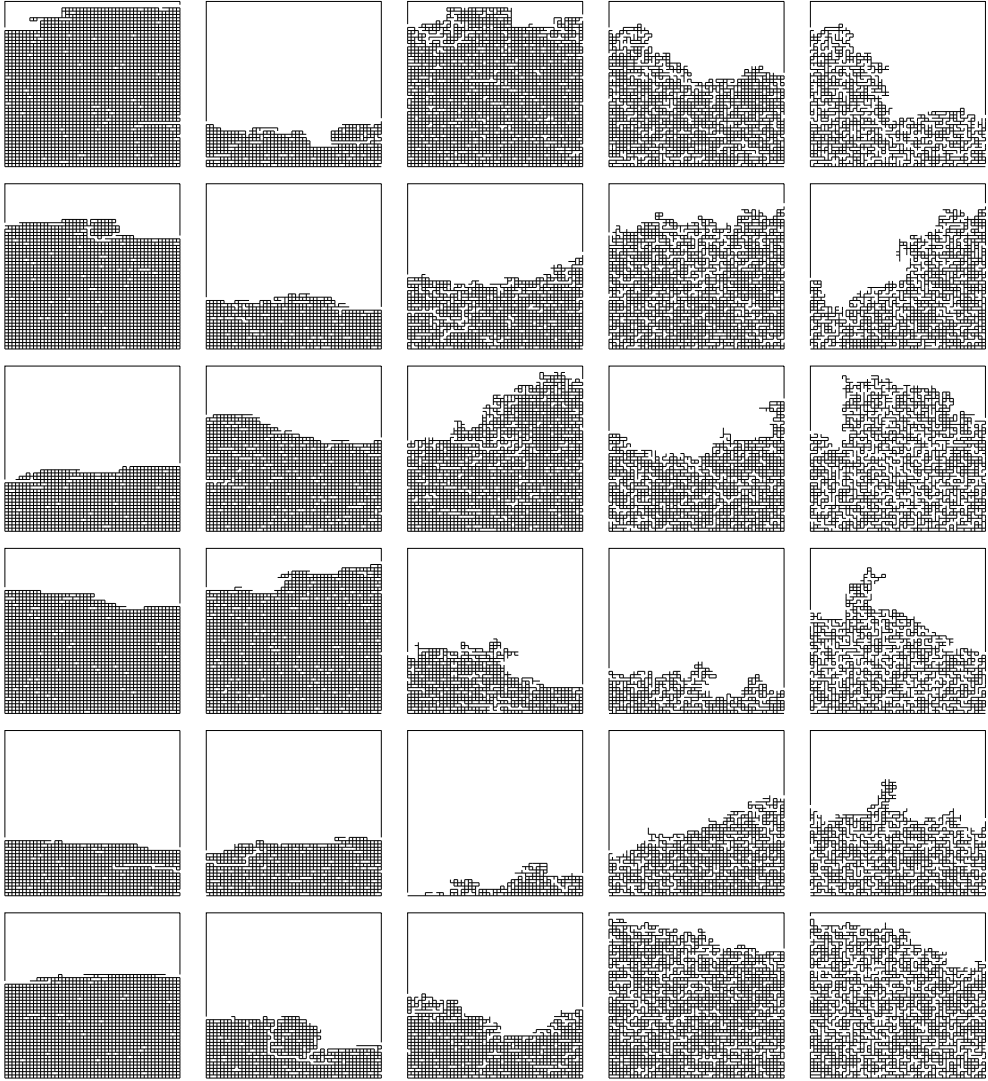}
\caption{The lower remaining part of a beam lattice of size 
         $L=50$ after it has
         been broken completely, shown for five different
         disorders, i.e., from left to right: $D=0.25$,
         $D=0.5$, $D=1$, $D=2$ and $D=4$, respectively.
         From top to bottom, six different samples have been 
         included in each case, the only difference 
         between the samples being the random casts generated 
         for the breaking thresholds.
         \label{i5x6}}
\end{figure*}
One of the reasons why lattice models are practical is the ease 
with which such disorder may be included. 

Presently we generate a 
random number $r$ on the unit interval $[0,1]$ and let this 
represent the cumulative threshold distribution. 
In Eq.~(\ref{bcri}) the breaking thresholds are now assigned as 
$t=r^{D}$, with
\begin{equation}
    p(t)=\frac{1}{D}t^{\frac{1}{D}-1}
              \label{diso}
\end{equation}
being the probability density. The same distribution is 
assumed for the threshold in axial force, $t=t_{F_{\rm C}}$, 
and bending moment, $t=t_{\mu{\rm C}}$, with the random casts, 
however, being different in the two cases.

There are now two types of distribution, i.e., $D>0$, in
which case
\begin{eqnarray}
    P(t)=t^{\frac{1}{D}}
\end{eqnarray}
is a cumulative distribution with bounds $0\le t<1$, and $D<0$, 
in which case
\begin{eqnarray}
    P(t)=1-t^{\frac{1}{D}}
\end{eqnarray}
is a cumulative distribution with bounds $1\le t<\infty$. In
this prescription $D=0$ corresponds to no disorder.
As~$|D|$ increases the coefficient of variation with respect to 
any two random numbers $r$ and $r^{\prime}$ on the interval~$[0,1]$ 
also increases, with the coefficients for $D>0$ and $D<0$ being 
reciprocal but otherwise the same. Hence, large values of $|D|$ 
correspond to strong disorders and small values to weak disorder. 
A few examples for $D>0$ have been included in Fig.~\ref{i5x6}, 
where the bottom part of the broken lattice is shown for five 
different disorders. Also shown is the sample-to-sample variation 
\begin{figure} [t] 
\includegraphics[angle=0,scale=0.63]{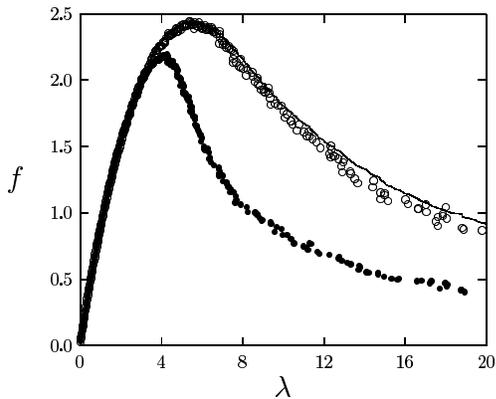}
\caption{Force $f$ versus displacement $\lambda$ for an $L=32$ 
         lattice with disorder $D=1$. The two upper curves 
         denote non-buckling fracture, with the continuous line 
         representing Eq.~(\ref{ma2x}) and open circles 
         representing Eq.~(\ref{ma3x}) with buckling suppressed. 
         The lower curve (filled circles) denotes buckling 
         fracture, i.e., the results of Eq.~(\ref{ma3x}) with the 
         out-of-plane degrees of freedom now included. 
         \label{L32ss}}
\end{figure}
for each of the disorders. For a given magnitude of $D$, 
the position of the final crack which breaks the system is seen to
vary from one sample to the next, as is its morphology. As the
disorder increases, so does the roughness of the crack
interface. The number of beams removed also increases with the
disorder, i.e., the $D=4$ samples are seen to be somewhat more
diluted than the $D=2$ samples. In Eq.~(\ref{diso}), $D>0$ and 
$D<0$ represent widely different types of distribution. While the 
former is a power law with a maximum threshold 
of one, and a tail which extends toward zero, the 
latter is a power law bounded below by a
minimum threshold of one, but now with a tail which extends
toward infinity. Both $D>0$ and $D<0$ are included in the 
present calculations. 

In the past many 
different distributions have been used to generate random breaking 
thresholds. However, as shown by Hansen et al.~\cite{hher}, as 
the system size diverges only the power law tails of the 
distribution, if they exist, towards zero or infinity should 
matter. Hence the use of $D$ as a parameter is very convenient, 
enabling the asymptotic behaviour of the fracture process to be 
fully explored as a function of the disorder.

\section{Stress and Strain}
\label{ss}
In the absence of structural disorder the crack now grows 
laterally from the site of the first beam removed, taking 
the shortest possible path across the lattice. Since in our
model the beams are linearly elastic up to the breaking
threshold, the first break triggers catastrophic rupture.
Stress and strain evolves differently in the presence of 
disorder. Now there are two 
competing mechanisms for crack growth. On the one hand, 
the presence of a crack causes stress to be intensified in 
its immediate vicinity, thereby lending bias towards the 
growth of already existing cracks. On the other hand, 
variations in material strength dictate that new cracks 
should instead appear in regions which are structurally 
weak. Which of the two mechanisms is the most important 
depends on the disorder regime. While in the case of 
strong disorder fracture is initially disorder dominated, 
it tends to be localized from the very beginning in the 
case of weak disorder. For strong disorders 
small cracks appear at random in the early stages of the 
process. Here the dominating feature is a wide distribution
of breaking thresholds. Since the weakest beams tend to
be removed first, the distribution gradually narrows as
more beams are removed. Simultaneously, with a growing 
number of cracks appearing on the lattice, a highly
non-uniform stress field develops. In other words, the
distribution of stresses widens. At the point where the
fracture process goes from being disorder dominated to
stress dominated, crack growth becomes localized~\cite{hher}. 
Smaller cracks now merge into a single dominating crack and 
the evolution of stress with strain goes from being stable 
to unstable. 

For a system of size $L=32$ and disorder $D=1$,
a comparison between the buckling and non-buckling stress-strain 
characteristics is shown in Fig.~\ref{L32ss}. 
The average stress and the average strain has been 
computed for every beam broken, and the number of samples 
involved is 10000 in the non-buckling case and 975 in the 
buckling case. Also included is the result of Eq.~(\ref{ma3x})
with the out-of-plane degrees of freedom suppressed.
This result is based on 525 numerical realizations. The
agreement between the non-buckling results of Eqs.~(\ref{ma2x}) 
and~(\ref{ma3x}) is seen
to be excellent, especially in the controlled and early
catastrophic regimes. Towards the end of the catastrophic
regime the loads obtained with Eq.~(\ref{ma3x}) are very
slightly lower than those that are obtained with Eq.~(\ref{ma2x}), a
result which can be ascribed to the presence of non-linear 
terms in the former. With buckling, a significant 
\begin{table}[b]
\caption{\label{tab1}
Ratio of buckling to non-buckling maxima, obtained for the 
external displacement $\lambda$ and force $f$, for disorder
$D=1$. The total number of samples calculated is $S$, and $L$ 
is the system size.
}
\begin{ruledtabular}
\begin{tabular}{cccrr}
  $L\hspace{7.5mm}$ & 
  $\lambda_{\rm Z}/\lambda_{0}$\footnote{Quantities labeled {\rm Z}
                                         refer to the buckling case.} & 
  $f_{\rm Z}/f_{0}$ & 
  $S_{\rm Z}$ & $S_{0}$ \\
\hline
 14$\hspace{7.5mm}$ & 0.83 & 0.93 & 1500 &  5000 \\
 17$\hspace{7.5mm}$ & 0.92 & 0.94 &  500 &  2500 \\
 20$\hspace{7.5mm}$ & 0.79 & 0.92 & 1000 &  1000 \\
 23$\hspace{7.5mm}$ & 0.83 & 0.92 &  203 &   800 \\
 27$\hspace{7.5mm}$ & 0.83 & 0.91 &  328 &   600 \\
 32$\hspace{7.5mm}$ & 0.75 & 0.90 &  975 & 10000 \\
 40$\hspace{7.5mm}$ & 0.74 & 0.89 &  210 &  1400 \\
 50$\hspace{7.5mm}$ & 0.77 & 0.91 &   70 &  1750 \\
 63$\hspace{7.5mm}$ & 0.77 & 0.89 &  110 &   700 \\
 80$\hspace{7.5mm}$ & 0.80 & 0.92 &   55 &   550 \\
\end{tabular}
\end{ruledtabular}
\end{table}
\begin{figure} [b]
\includegraphics[angle=0,scale=0.72]{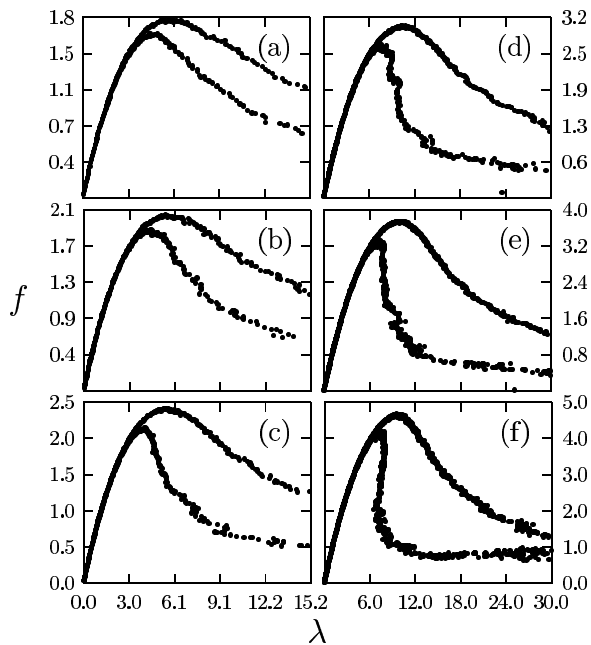}
\caption{Force $f$ versus displacement $\lambda$ for 
         a range of system sizes with disorder $D=1$, i.e.,
         for (a)~$L=23$, (b)~$L=27$, (c)~$L=32$, (d)~$L=40$,
         (e)~$L=50$ and (f)~$L=63$. In each case the top
         curve is the non-buckling result of the simple beam 
         model, calculated from Eq.~(\ref{ma2x}), and the curve 
         below is the buckling result, calculated from 
         Eq.~(\ref{ma3x}). The labels on the axes are scaled 
         down from those in plot~(f), being otherwise 
         proportional to system size.
         \label{6plot}}
\end{figure}
\begin{figure} [t] 
\includegraphics[angle=0,scale=0.63]{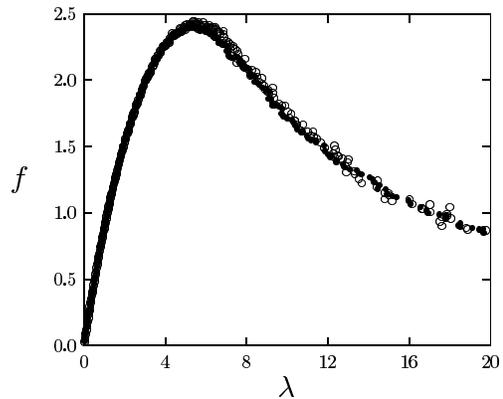}
\caption{Force $f$ versus displacement $\lambda$ for an $L=32$ 
         lattice with disorder $D=1$, comparing the buckling 
         (filled circles) and the non-buckling (open circles) 
         results of Eq.~(\ref{ma3x}) where, 
         in the former case, $\chi=0$ in Eq.~(\ref{ft}).
         \label{supL32ss}}
\end{figure}
reduction is obtained in both maximum strength and displacement. 
There is also a notable difference in the shape of 
the curve within the catastrophic regime. Here the 
response is less stable with respect to displacement control. 
That is, 
the force in the catastrophic regime
falls off more rapidly as the displacement increases. 

Stress and strain for a range of system sizes is shown in
Fig.~\ref{6plot} for the same disorder, i.e., $D=1$. In 
calculations for the non-buckling beam lattice, involving a
much larger range of sizes~\cite{tbp2}, the scaling 
with $L$ of the top of the stress-strain curve is found to 
be characterized by an exponent close to unity. The 
stress-strain curves can then be made to collapse onto 
each other by scaling the axes according to
\begin{eqnarray}
    f/L^{\gamma}=\phi(\lambda/L^{\gamma}),
                  \label{slaw}
\end{eqnarray}
where $\gamma\approx 1$ and $\phi$ is a scaling function. 
Since there is no reason why the buckling system should 
behave according to different laws in this respect, the
reduction in stress and strain should itself be proportional
to system size. As was noted in section~\ref{planb}, 
fracture is initiated by imposing on the top row of the 
lattice a displacement of one beam length. Hence, to avoid 
scale effects on the buckling behaviour, one of the factors 
$L$ introduced in the stress enhancement factor, i.e., in 
Eq.~(\ref{tsens}), is a scale factor.
Without this factor, a different value of the exponent 
$\gamma$ would be obtained in Eq.~(\ref{slaw}). With the
current choice of parameters, maximum stress and strain
in the buckling and non-buckling cases scale according 
to the same law, as can be seen from Fig.~\ref{6plot}, 
and also from the comparison of buckling and non-buckling 
stress-strain maxima in Table~\ref{tab1}. Here the values 
obtained for the reduction in maximum stress and strain
appear to be
consistent for systems larger than about $L=20$. Below
this, finite size effects become apparent. The most
reliable estimates are obtained with the largest number
of calculated samples, hence, for $D=1$, buckling reduces
the maximum strength by about $10$\% and the maximum 
displacement by about $25$\%.
The shape of the curve in the catastrophic
regime varies according to system size. Beyond
the turn-over point between stable and unstable crack
growth, rupture in the non-buckling system is seen to become 
increasingly less stable as the size of the system
increases. This effect is even more pronounced when the
sheet is allowed to buckle.

Significant differences are evident in a comparison between
the force or displacement fields of, say, a uniform center-cracked 
lattice in the case of buckling with the corresponding force or 
displacement fields in the non-buckling case. Specifically, the
transverse forces near to the crack edges, which are compressive 
in the non-buckling lattice, are released when the lattice
buckles, causing the flanks of the crack to deflect. Since the 
alterations in the force or displacement fields extend beyond the
immediate neighbourhood of the crack tips, one may ask whether
these effects in themselves are sufficient to bring about a
reduction in the maximum load carrying capacity of the lattice.
The mechanism by which the stress is intensified at the crack
tips, however, takes place on a scale smaller than the individual 
beam, which is why the fracture criterion Eq.~(\ref{bcri}) has 
been augmented by the factor $\chi$. Hence, the 
hypothetical case of fracture where buckling does not induce 
intensified stress at the crack tips can be investigated simply 
\begin{figure} [t]
\includegraphics[angle=0,scale=0.72]{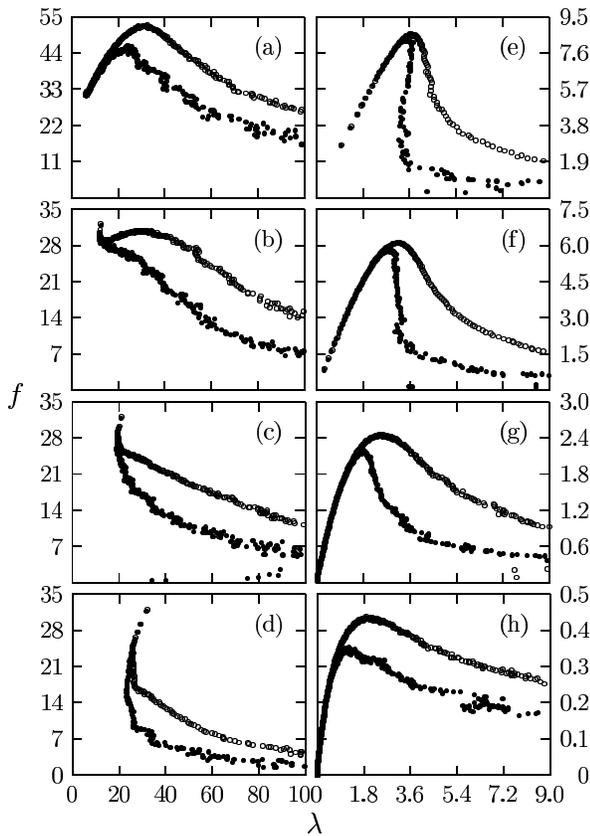}
\caption{Force $f$ versus displacement $\lambda$ obtained for a 
         lattice system of size $L=32$, for a range of disorders 
         with both $D<0$ and $D>0$. On the left are shown (a)~$D=-5$, 
         (b)~$D=-3$, (c)~$D=-2$ and (d)~$D=-1$. On the right are 
         shown (e)~$D=0.33$, (f)~$D=0.5$, (g)~$D=1$ and (h)~$D=2$. 
         In (a)--(c) the respective scales on $\lambda$ are $\times5$, 
         $\times2.5$ and~$\times1.5$ that in~(d), and in~(e)--(g) 
         the respective ratios $f/\lambda$ are 0.38, 0.3, and~0.15.
         In all cases the stress-strain curve for buckling lies 
         below that which does not buckle.
         \label{alld}}
\end{figure}
by setting $\chi=0$ in Eq.~(\ref{ft}).  In Fig.~\ref{supL32ss} 
the result is compared with that obtained in non-buckling 
fracture. Clearly, the evolution of stress with strain is seen to 
be the same in both the stable and catastrophic regimes. Both 
curves were obtained from Eq.~(\ref{ma3x}), based on 1350 
samples in the buckling case and 525 samples in the non-buckling 
case. One can thus state with certainty that it is the intensified
stress at the crack tips, due to a coupling between in-plane
and out-of-plane deformations, rather than the re-distribution 
of stresses within the buckling zone (but away from the immediate
neighbourhood of the crack tips) which causes the reduction in 
residual strength. 

In Fig~\ref{alld} is shown the effect disorder has on the 
interaction of buckling with fracture.
Plots on the left-hand side display stress-strain
curves for $D<0$ type disorder.
Here the initial response, i.e., the linear
relationship which extends from the origin to the data point
of the first broken beam, has been omitted. There is
no tail towards zero in the distribution of thresholds here,
and consequently
there are no broken beams on this part of the curve. 
In plot~(d), where $D=-1$, the first beam to break triggers 
a catastrophic rupture mode
which backtracks along the initial linear response
for the first few breaks. It then encounters a vertical section 
of the curve where several values of $f$ correspond to the
same $\lambda$. If displacement control is applied and relaxed
sufficiently fast for the crack to be halted at this point, 
we have a situation of conditional stability where a slight 
perturbation, say a bump or a jar, suffices to further propagate 
the crack (this refers to the average situation, with
individual samples being subject to fluctuations). 
Otherwise, applying displacement control without this sudden
relaxation, the crack develops catastrophically until it is 
arrested when encountering strong beams in the tail towards 
infinity. From here on, the force continues to fall off as the
displacement is increased. 

The main effect buckling has for weak $D<0$ disorder
is to make the force fall off more rapidly in the
catastrophic regime. Additionally, the section of the curve 
which is conditionally stable in the non-buckling case
is rendered unstable, i.e., the
curve turns back on itself.
As $|D|$ increases there is a turn-over in the average 
stress-strain behaviour in the sense that, beyond $D=-3$,
force control may be applied without necessarily triggering
catastrophic rupture. This, of course, is due to the presence
of a large number of beams with high breaking thresholds.
When the tail towards infinity 
becomes sufficiently strong, in other words, the number
of beams which can be found in the vicinity of the lower
bound becomes a minority. The stress-strain relationship
then attains a similar form to that of $D>0$, except now
fracture starts at a finite displacement or force. Although
in (a) the controlled regime, which obtains after the first 
beam breaks, contains a smaller number of broken beams than
does, for instance, the one in (g), the reduction in
force and displacement due to buckling is comparable in
the two cases. The reason for this is a more intense stress
field in the former case, caused by higher thresholds, which 
in turn moves the onset of buckling to an earlier stage
of the fracture process.

\begin{table}
\caption{\label{tab2}
Ratio of buckling to non-buckling maxima, obtained for the 
displacement $\lambda$ and force $f$, for
$L=32$. The number of samples is $S$, and $D$ 
is the disorder.
}
\begin{ruledtabular}
\begin{tabular}{cllrr}
  $D\hspace{7.5mm}$ & 
  $\lambda_{\rm Z}/\lambda_{0}$\footnote{Quantities labeled {\rm Z}
                                         refer to the buckling case.} & 
  $f_{\rm Z}/f_{0}$ & 
  $S_{\rm Z}$ & 
  $S_{0}$ \\
\hline
  0.2$\hspace{7.5mm}$ & 1.   & 1.  $\hspace{7.5mm}$ &  200 &   500 \\
0.333$\hspace{7.5mm}$ & 0.98 & 0.97$\hspace{7.5mm}$ &  200 &   500 \\
  0.5$\hspace{7.5mm}$ & 0.92 & 0.96$\hspace{7.5mm}$ &  280 &  3500 \\
    1$\hspace{7.5mm}$ & 0.75 & 0.90$\hspace{7.5mm}$ &  975 & 10000 \\
    2$\hspace{7.5mm}$ & 0.65 & 0.81$\hspace{7.5mm}$ &  192 &  1000 \\
\end{tabular}
\end{ruledtabular}
\end{table}
Displayed on the right-hand side are stress-strain
curves with $D>0$ type disorder. These
are mostly subject to the same features as the result of 
Fig.~\ref{L32ss}, relevant to $D=1$. An exception, perhaps,
is $D=2$, for which the stability in the catastrophic 
regime appears to be unchanged by buckling. For $D=2$ and
beyond, however, the number of beams relevant to the
catastrophic regime is small compared to that of the
controlled regime. This means that the number of samples which
contribute decreases toward the end of the stress-strain
curve (the curves have not been truncated at the average
number of broken beams), and hence statistical fluctuations 
become large in this region. 

Whereas for weak $D>0$ 
disorders only a small reduction is obtained in the 
maximum of stress and strain, the stability in the 
catastrophic regime of fracture is significantly affected
in this case, as can be seen from (e) in Fig.~\ref{alld}, 
i.e., for $D=0.33$. The reason is the onset of 
buckling, which for low disorders occurs near the top of
the curve. Even when the disorder is sufficiently low
for the onset of buckling, in average, to occur after the 
top has been reached, a slight decrease in maximum strength 
may be expected. This, of course, is due to the fact that 
a number of samples will buckle prior to this average onset.
 
In Table~\ref{tab2}, results for the $L=32$ system are 
shown for a range of disorders with $D>0$. Here the decrease 
in force and displacement is seen to depend on the magnitude 
of $D$, i.e., as $|D|$ increases buckling has
an increasingly adverse effect
on both the maximum load and the maximum displacement a
disordered system can sustain. The maximum displacement is
more strongly affected than the maximum load.

\section{Summary}
\label{ccon}
The breaking characteristics of thin sheets with
structural disorder have been obtained in numerical 
simulations which include the out-of-plane buckling behaviour. 
The model used is an elastic lattice of beams
where each beam is representative of the scale of the structural
disorder. Depending on the magnitude of disorder, breakdown
is either localized to the first point of
damage or initially a random cracking process 
which at a later stage crosses over to localized fracture 
behaviour. 

The breakdown process is initiated from an
initially intact sheet, where buckling sets in after a
certain amount of damage has occurred. Specifically,
the onset of buckling varies
considerably according to both the size and configuration of
the emerging cracks. 
Given a certain system size and
disorder, several numerical realizations of a sheet are 
generated, corresponding to different sets of random
breaking thresholds.
The statistical properties are then
obtained from the average behaviour based on the
disorders and sizes chosen.

As in the case of uniform pre-cracked sheets, it is found
that buckling adversely affects the external
force and displacement a randomly disordered sheet
can sustain in mode-I type tensile loading. 
The degree to which the maximum force
and displacement is reduced depends on the
magnitude of the disorder. For instance, in a material such 
as paper this would mean that buckling should affect the 
maximum load carrying capacity more adversely in the case of 
a fibre-web with uneven formation than one with a more
even formation. 
When the meso-scale disorder is low 
the reduction in strength is insignificant and it is the 
catastrophic regime which is most affected, now being less 
stable. 

\end{document}